\documentclass[12pt]{article}
\usepackage{hyperref,amssymb,amsmath,graphicx,verbatim,amsthm}
\usepackage{color}
\newcommand{\ignore}[1]{} 

\newtheorem{thm}{Theorem}
\newtheorem{cor}[thm]{Corollary}
\newtheorem{lem}[thm]{Lemma}
\newtheorem{prop}[thm]{Proposition}
\newtheorem{clm}[thm]{Claim}

\newtheorem*{thm*}{Theorem}

\newcommand\minvar{{\hbox{minv}}}

\theoremstyle{definition}

\theoremstyle{remark}

\numberwithin{equation}{section}

\newcommand{\set}[1]{\left\{#1\right\}}

\newcommand{\N}{\mathbb{N}}

\newcommand{\R}{\mathbb{R}}

\newcommand{\eps}{\varepsilon}

\def\squareforqed{\hbox{\rlap{$\sqcap$}$\sqcup$}}
\def\qed{\ifmmode\squareforqed\else{\unskip\nobreak\hfil
\penalty50\hskip1em\null\nobreak\hfil\squareforqed
\parfillskip=0pt\finalhyphendemerits=0\endgraf}\fi}

\newcommand{\F}{\mathbb{F}}

\newcommand{\wh}[1]{\widehat{#1}}

\begin{document}

\title{Homogeneous formulas and symmetric polynomials
\author{
Pavel Hrube\v{s}\thanks{School of Mathematics, Institute for Advanced Study,
 Princeton NJ. Emails:~\texttt{pahrubes@centrum.cz} and \texttt{amir.yehudayoff@gmail.com}.
 Partially supported by NSF grant CCF 0832797.} \and
Amir Yehudayoff\footnotemark[1]}
}

\date{ }

\maketitle

\begin{abstract}
We investigate the arithmetic formula complexity of the elementary symmetric polynomials $S^k_n$.
We show that every multilinear homogeneous formula computing $S^k_n$ has size at least $k^{\Omega(\log k)}n$,
and that product-depth $d$ multilinear homogeneous formulas for $S^k_n$ have size at least $2^{\Omega(k^{1/d})}n$.
Since $S^{n}_{2n}$ has a multilinear formula of size $O(n^2)$,
we obtain a superpolynomial separation between multilinear and multilinear homogeneous formulas.
We also show that $S^k_n$ can be computed by homogeneous formulas of size $k^{O(\log k)}n$,
answering a question of Nisan and Wigderson.
Finally, we present a superpolynomial separation between monotone and non-monotone formulas
in the noncommutative setting, answering a question of Nisan.
\end{abstract}


\newpage

\section{Introduction}

We address two basic topics in arithmetic complexity:
the power of homogeneity and computing the symmetric polynomials.
A basic structural result \cite{Hyafil,VSBR} in arithmetic complexity asserts that

\smallskip
\emph{
$(\star)$ \ \ if a homogeneous polynomial of degree $k$ has a formula of size $s$,
then it has a homogeneous formula of size at most $s^{O(\log k)}$. }

\smallskip

A natural question to ask is whether the upper bound given by $(\star)$ is tight,
or perhaps the true bound is, say, $O(k s)$?
With our current techniques this question is unfortunately out of reach.
Most importantly, superpolynomial lower bounds on homogeneous formula complexity (for low degree polynomials) are not known.
Still, we can investigate this question in restricted models of computation;
we investigate the multilinear setting.

The elementary symmetric polynomials $S^k_n$ (formally defined below)
seem to be good candidates for a separation in $(\star)$.
Over an infinite field, they have non-homogeneous formulas of size $O(n^2)$.
In \cite{NW} it was conjectured that $S^k_n$ requires
homogeneous formulas of size at least $n^{\Omega(\log k)}$,
matching the upper bound given in $(\star)$. This, however, is not the case --
we show that $S^k_n$ has homogeneous formulas of size $k^{O(\log k)}n$, which is linear for a fixed $k$.
In fact, the conjecture does not even hold for monotone formulas --
$S^k_n$ have monotone formulas of size $n^{1+o(1)}$, if $k$ is fixed.

The main part of this paper is devoted to multilinear homogeneous formulas computing $S^k_n$.
In particular, we show that $S^{n}_{2n}$ requires homogeneous multilinear formulas of superpolynomial size.
This implies a superpolynomial separation between multilinear and homogeneous multilinear formulas.
However, this bound does not match the bound given by $(\star)$ --
for a general $k$, the lower bound is of the form $k^{\Omega(\log k)}n$
(rather than $n^{\Omega(\log k)}$).

\subsection{Results}

Let us first give the usual definitions.
An \emph{arithmetic circuit} $\Phi$ over the field $\F$ is a directed acyclic graph as follows.
Every node in $\Phi$ of in-degree $0$ is labelled by either a variable or a field element in $\F$.
Every other node in $\Phi$ has fan-in at least two and is labelled by either $\times$ or $+$.
Nodes labelled by $\times$ are \emph{product nodes}, and nodes labelled by $+$ are \emph{sum nodes}.
An arithmetic circuit is called a \emph{formula},
if the out-degree of every node in it is one.
A circuit $\Phi$ computes a polynomial $\wh{\Phi}$ in the obvious manner.

A polynomial $f$ is \emph{homogeneous} if the total degrees
of all the monomials that occur in $f$ are the same.
A polynomial $f$ is \emph{multilinear} if the
degree of each variable in $f$ is at most one.
A circuit $\Phi$ is \emph{homogeneous} if every node in $\Phi$ computes a homogeneous polynomial.
A circuit $\Phi$ is \emph{multilinear} if every node in it computes a multilinear polynomial.
A circuit $\Phi$ over the real numbers is called \emph{monotone}
if every field element in $\Phi$ is a nonnegative real number.

We define the \emph{size} of a formula as the number of leaves in it\footnote{The 
total number of nodes in a tree where each internal node has fan-in at least two is at most twice the number of leaves.}.
The \emph{depth} of a formula is the length of the longest directed path in it.
The \emph{product-depth} of a formula $\Phi$ 
is the largest number of product nodes in a directed path in $\Phi$.

The elementary symmetric polynomial $S^k_n$ is the polynomial in variables $x_1,\ldots,x_n$ defined as
\[\sum_{ i_1< i_2<\cdots <i_k} x_{i_1} x_{i_2} \cdots x_{i_k} ; \]
it is a homogeneous multilinear polynomial of degree $k$.

We show the following lower bounds on the size of multilinear homogeneous formulas computing $S^k_n$.

\begin{thm}
\label{thm: lb}
Let $n\geq 2k$ and $d$ be nonzero natural numbers.
\begin{enumerate}
\item Every homogeneous multilinear formula computing $S^k_n$ has size at least $k^{\Omega(\log k)} n$.
\item Every homogeneous multilinear formula of product-depth $d$ computing $S^k_n$ has size at least
$2^{\Omega(k^{1/d})}n$.
\end{enumerate}
\end{thm}

In the case of $S^n_{2n}$, the first lower  bound is superpolynomial and the latter exponential.
Since the symmetric polynomials have multilinear formulas of size  $O(n^2)$ and product-depth one
(see Section~\ref{sec: Ben-Or}),
the theorem shows that homogeneous multilinear formulas
are superpolynomially weaker than multilinear formulas,
and that constant depth homogeneous multilinear formulas are exponentially weaker
than their nonhomogeneous counterparts.
Since monotone formulas are both homogeneous and multilinear,
we have a superpolynomial separation between monotone and non-monotone formulas.
This separation also holds in the noncommutative case,
which answers a question raised in \cite{Nisan}.
The lower bounds are based on counting the number of monomials that occur
in a polynomial that is computed by a homogeneous multilinear formula.
We get essentially the same bounds as \cite{ShamirSnir} get in the case of monotone
formulas\footnote{Our first lower bound can also be viewed as corollary of the bound in \cite{ShamirSnir}.}.
However, our arguments are different than in \cite{ShamirSnir},
which may be useful for more general cases as well. 

We also provide upper bounds on the formula complexity of $S^k_n$.

\begin{thm}
\label{thm: ub}
Let $n,k$ be nonzero natural numbers.
\begin{enumerate}
\item $S^k_n$ has a homogeneous formula of size $k^{O(\log k)} n$.
\item $S^k_n$ has depth four (product-depth two) homogenous formula of size $2^{O(k^{1/2})} n$.
\item $S^k_n$ has a monotone formula of size \[2n\cdot n^{\log\left(\frac{k-1}{\log (2n)}+1\right)}\cdot \left (\frac{\log (2n)}{k-1}+1\right)^{k-1} = n^{O(\log(\frac{k}{\log n}))}.\]
\end{enumerate}
\end{thm}

For fixed $k$, all of the upper bounds given by Theorem~\ref{thm: ub} are essentially linear in $n$ (i.e., linear in the first two cases, and $n^{1+o(1)}$ in the last one).

\section{Lower bounds}

In this section we prove the lower bounds given by Theorem~\ref{thm: lb}.

\subsection{Technical estimates}

We need the following technical estimate.

\begin{lem}
\label{lem: new binomial}
Let $n \geq 2k$ be nonzero natural numbers.
Fix nozero natural numbers $k_1,\ldots,k_p$ such that $k_1 + \cdots + k_p = k$.
Then for every natural number $n_1,\ldots,n_p$ such that $n_1 + \cdots + n_p = n$,
\begin{align*}
{{n_1}\choose{k_1}} \cdots {{n_p}\choose{k_p}}
\leq 3 k^{1/2} (k_1 \cdots k_p)^{-1/2} {n \choose k} .
\end{align*}
\end{lem}

\begin{proof} 
1) We shall first prove the lemma using the additional assumption that $k_i\geq 2$ for every $i=1,\ldots,p$. 
We estimate the maximum of 
${{n_1}\choose{k_1}} \cdots {{n_p}\choose{k_p}}$ with respect to $n_1,\ldots,n_p$ satisfying the given constraints.

First we show that we can assume $1.5 k_i \leq n_i$ for every $i \in [p]$.
Let $n_1,\ldots,n_p$ be the integers where the maximum is attained.
Assume without loss of generality that $n_1/k_1 \geq n/k \geq 2$.
For every $i \in \set{2,\ldots,p}$,
the choice of $n_1,\ldots,n_p$ implies that
${n_1 - 1 \choose k_1} {n_i+1 \choose k_i} \leq {n_1 \choose k_1} {n_i \choose k_i}$.
Hence $(n_i + 1)/(n_i + 1 - k_i) \leq n_1 / (n_1 - k_1)$,
and so $n_i / k_i \geq n_1 / k_1 - 1 / k_i \geq 2 - 1/2$.

For $i=1,\ldots,p$ and a real number $z$ such that $z > k_i$, define
$f_i(z) = \frac{z^z}{k_i^{k_i} (z-k_i)^{z-k_i}}$.
Thus $\frac{\partial}{\partial z} f_i = f_i \cdot \ln (1/(1-k_i/z))$.
Denote
\begin{align*}
F(z_1,\ldots,z_p) = f_1(z_1) f_2(z_2)  \cdots f_p(z_p) .
\end{align*}
We shall determine the maximum of $F$ on the set $S \subset \R^p$ defined
by the constraints $z_1 + \cdots + z_p = n$ and $z_i \geq 1.5 k_i$, $i = 1,\ldots,p$. 
Since $S$ is compact and $F$ continuous, $F$ has a maximum on $S$.
Let $(z_1,\ldots,z_p) \in S$ be the point at which $F$ attains its maximum.
Our goal is to show that $z_i/k_i = n/k$ for every $i = 1,\ldots,p$.
Assume without loss of generality that $z_1/k_1 \leq z_i/k_i$ for every $i = 2,\ldots,p$.
Assume towards a contradiction that there exists $i = 2,\ldots,p$ with $z_1/k_1 < z_i/k_i$, 
and consider $f(z_1+x)f(z_i-x)$ as a function of $x$.
Since 
$$\frac{\partial}{\partial x} f(z_1+x)f(z_i-x) \Big|_{x = 0} = 
f(z_1)f(z_i) \ln \left( \frac{1-k_i/z_i}{1-k_1/z_1} \right) > 0,$$
there exists $\eps > 0$ such that $(z_1 + \eps,\ldots,z_i- \eps,\ldots,z_p) \in S$ and
$f(z_1+\eps)f(z_i-\eps) > f(z_1)f(z_i)$; a contradiction to the choice of $z_1,\ldots,z_p$. 
Hence, since $z_1 + \cdots + z_p = n$ and $k_1 + \cdots + k_p = k$,
we have $z_i/k_i = n/k$ for every $i = 1,\ldots,p$. 
So the maximum value of $F$ on $S$ is
\begin{align*}
 \prod_{i=1,\ldots,p}  \frac{n^{k_i}}{ k^{k_i}}
\frac{n^{n-k_i}}{(n-k)^{z_i-k_i}}=\frac{n^{n}}{ k^{k} (n-k)^{n-k}}
\end{align*}
Stirling's approximation tells us that for every nonzero $N,K \in \N$ with $1.5K \leq N$,
\begin{align*}
(1/3) K^{-1/2} \frac{N^{N}}{ K^{K} (N-K)^{N-K}} \leq {N \choose K}
\leq  K^{-1/2} \frac{N^{N}}{ K^{K} (N-K)^{N-K}},
\end{align*}
which implies
\begin{align*}
{n_1 \choose k_1} \cdots {n_p \choose k_p} & \leq (k_1 \cdots k_p)^{-1/2} F(n_1,\dots n_p)
\\ & \leq (k_1 \cdots k_p)^{-1/2} \frac{n^n}{k^k(n-k)^{n-k}} 
\\ & \leq 3 k^{1/2} (k_1 \cdots k_p)^{-1/2} {n \choose k} .
\end{align*}

2) Assume without loss of generality that $k_1,\ldots,k_\ell = 1$,
and denote $k' = k_1 + \cdots + k_\ell$ and $n' = n_1 + \cdots + n_\ell$.
Since 
${{n_1}\choose{k_1}} \cdots {{n_\ell}\choose{k_\ell}} \leq {n' \choose k'}$,
part 1) shows   that
\begin{align*}{{n_1}\choose{k_1}} \cdots {{n_p}\choose{k_p}} 
&\leq& 3 k^{1/2} (k_{\ell+1} \cdots k_p)^{-1/2} {n - n' \choose k - k'} {n' \choose k'} 
\leq 3 k^{1/2} (k_1 \cdots k_p)^{-1/2} {n \choose k} . 
\end{align*}
\end{proof}

\subsection{In-degree two}
\label{sec: formulas}

Let $f$ be a homogeneous polynomial of degree $k$.
We say that $f$ is \emph{balanced} if there exist $p$ homogeneous polynomials
$f_1,\ldots,f_p$ such that $f = f_1 f_2 \cdots f_p$ with
\begin{enumerate}
\item $(1/3)^i k < \deg f_i\leq (2/3)^i k$, $i=1,\ldots,p-1$, and
\item $\deg(f_p)=1$ .
\end{enumerate}
For a balanced polynomial $f$, denote by $\minvar(f)$ the number of variables that occur in $f_p$.

The following lemma shows that a small homogeneous formula
can be written as a short sum of balanced polynomials.

\begin{lem}
\label{lem: structure of formulas}
Let $\Phi$ be a homogeneous formula with in-degree at most two of size $s$ and degree $k > 0$.
Then there exist balanced polynomials $f_1,\ldots,f_{s^\prime}$ such that $s^\prime\leq s$, 
$$\wh{\Phi} = f_1 + \cdots + f_{s^\prime}$$
and $\sum_{i=1,\ldots , s} \minvar(f_i) \leq s$.
If $\Phi$ is multilinear, so are $f_1,\ldots,f_{s^\prime}$.
\end{lem}

For a node $w$ in a formula $\Phi$, denote by $\Phi_w$ the sub-formula of $\Phi$
with output node $w$, and
by $\Phi_{(w=\alpha)}$ the formula obtained by deleting the edges going into $w$
and labeling $w$ (which is now an input node) by the field element $\alpha$.

\begin{proof}
Let us first note the following:
\begin{clm}
\label{clm: exists w in formula}
If $\Phi$ is a formula of degree $k \geq 2$, then there exists
a node $w$ in $\Phi$ such that $(1/3)k \leq \deg(w)< (2/3) k$,
where $\deg(w) = \deg(\wh{\Phi}_w)$.
\end{clm}

\begin{proof}
There exists a node $v$ in $\Phi$ such that $\deg(v) \geq (2/3)k$,
but for every child $w$ of $v$ (i.e., the edge $(w,v)$ occurs in $\Phi$), $deg(w) < (2/3)k$.
Hence $v$ is a product node $v=w_1 \times w_2$.
If $\deg(w_1) \geq \deg(w_2)$ then $w=w_1$ has the correct properties, otherwise set $w=w_2$.
\end{proof}

We prove the lemma by induction on $s$ and $k$.
If $k=1$, $\wh{\Phi}$ is a balanced polynomial and $\minvar(\wh{\Phi}) \leq s$,
since $\Phi$ contains at most $s$ variables.
Assume that $k \geq 2$.
Let $w$ be a node in $\Phi$ of degree $k^\prime$ such that $(1/3)k\leq k^\prime<(2/3)k$;
the node $w$ exists by Claim~\ref{clm: exists w in formula}.
Homogeneity implies that we can write
$$\wh{\Phi} = h \cdot \wh{\Phi}_w + \wh{\Phi}_{(w=0)},$$
where $h$ is a polynomial of degree $k-k^\prime$.
Let $s_w$ denote the size of $\Phi_w$ and let $s_{(w=0)}$ denote the size of $\Phi_{(w=0)}$.
Thus $s_w + s_{(w=0)} \leq  s$.
By the inductive assumption,
$\wh{\Phi}_w = h_1+ \cdots + h_{s_w^\prime}$ and
$\wh{\Phi}_{w=0} = g_1+ \cdots + g_{s_{(w=0)}^\prime}$,
where $s_w^\prime\leq s_w$, $s_{w=0}^\prime\leq s_{w=0}$, 
$h_1,\ldots,h_{s_w^\prime}$ are balanced polynomials such that $\sum_i \minvar(h_i) \leq s_w$,
and $g_1,\ldots,g_{s_{(w=0)}^\prime}$ are balanced polynomials such that $\sum_j \minvar(g_j) \leq s_{(w=0)}$.
(It may happen that $\wh{\Phi}_{(w=0)}$ is the zero polynomial.)
Hence
\begin{equation}\label{eq:balanced}
\wh{\Phi}= h h_1+ \cdots + hh_{s_w} + g_1 + \cdots + g_{s_{(w=0)}}.
\end{equation}
Since $(1/3)k<\deg h\leq (2/3)k$ and $(1/3)k\leq k^\prime<(2/3)k$, $h h_i$ is a balanced polynomial of degree $k$.
Hence (\ref{eq:balanced}) is an expression of $\wh{\Phi}$ in terms of balanced polynomials.
Moreover, $\minvar(hh_i) = \minvar(h_i)$, and hence
$\sum_i \minvar(h h_i) + \sum_j \minvar(g_j) \leq s_w + s_{(w=0)} \leq s$.

In the case that $\Phi$ is multilinear,
we can assume without loss of generality that $\Phi$ is in fact syntactically multilinear
(see, for example, \cite{R04a}),
that is, for every product node $v = v_1 \times v_2$ in $\Phi$,
the set of variables that occur in $\Phi_{v_1}$ and the set of variables
that occur in $\Phi_{v_2}$ are disjoint.
This implies that the polynomials $hh_1,\ldots,h h_{s'_w}$ are multilinear.
The lemma follows by induction.
\end{proof}

The following lemma bounds the number of monomials in a balanced polynomial.

\begin{lem}
\label{lem: number monomials in balanced mult}
Let $f$ be a balanced multilinear polynomial of degree $k$ with at most $n$ variables, $2k \leq n$.
Then the number of monomials that occur in $f$ is at most
$$3 k^{- c \log k + 3/2} {n \choose k} \minvar(f) / n ,$$
where $c>0$ is a universal constant.
\end{lem}

\begin{proof}
Assume that $f=f_1 \cdots f_p$, where $f_i$ has degree $k_i$ and $n_i$ variables
(so $n_p = \minvar(f)$).
Homogeneity implies $k_1 + \cdots + k_p = k$ and multilinearity implies $n_1 + \cdots + n_p \leq n$ 
(without loss of generality we can assume that $n_1 + \cdots + n_p = n$). 
Since each $f_i$ is also homogeneous and multilinear,
it contains at most $\binom{n_i}{k_i}$ monomials.
Thus, since $k_p = 1$, $f$ contains at most
${n_1 \choose k_1} \cdots \binom{n_{p-1}}{k_{p-1}} n_p$ monomials,
which, by Lemma~\ref{lem: new binomial}, is at most
$3 k^{1/2} (k_1 \cdots k_p)^{-1/2} {n - n_p \choose k - 1} n_p$.
For every $1 \leq i \leq \log k / (2 \log 3)$, we have $k_i \geq k^{1/2}$, and so
$3 (k_1 \cdots k_p)^{-1/2} \leq 3k^{- c \log k}$ with $c > 0$ a universal constant.
Since $\binom{n-n_p}{k-1}\leq \binom{n-1}{k-1}=\binom{n}{k}\frac{k}{n} $, 
the number of monomials that occur in $f$ is at most
\begin{align*}
3k^{-c \log k+1/2} {n-n_p \choose k - 1} n_p \leq 3 k^{- c \log k + 3/2} {n \choose k} \frac{ \minvar(f) }{ n } .
\end{align*}
\end{proof}

We can now bound the number of monomials in a polynomial by its multilinear homogeneous formula complexity.

\begin{prop}
\label{prop: ub num of monomials indeg two}
Let $\Phi$ be a multilinear homogeneous formula with in-degree at most two. Assume that $\Phi$ has 
size $s$, degree $k>0$ and at most $n$ variables, $2k \leq n$.
Then the number of monomials that occur in $\wh{\Phi}$ is at most
\[3 k^{- c \log k + 3/2} {n \choose k} \frac{s}{ n},\]
where $c$ is a universal constant.
\end{prop}

\begin{proof}
By Lemma~\ref{lem: structure of formulas},
there exist balanced multilinear polynomials $f_1,\ldots,f_{s^\prime}$ such that
$\wh{\Phi} = f_1 + \cdots + f_{s^\prime}$ and $\sum_{i=1,\ldots , s^\prime} \minvar(f_i) \leq s$.
By Lemma~\ref{lem: number monomials in balanced mult},
there exists a constant $c > 0$ such that for every $i = 1,\ldots,s^\prime$,
the number of monomials that occur in $f_i$ is at most
$3 k^{- c \log k + 3/2} {n \choose k} \minvar(f) / n $.
The proposition follows, since the number of monomials that occur in $\wh{\Phi}$
is at most the sum of the number of monomials that occur in the $f_i$'s.
\end{proof}

\begin{cor} The first part of Theorem~\ref{thm: lb} holds.
\end{cor}

\begin{proof}
The number of monomials in $S^k_n$ is ${n \choose k}$.
\end{proof}

\subsection{Bounded depth}

A homogeneous polynomial $f$ has a $(p,\ell)$-\emph{form}
if there exist homogeneous polynomials $f_1,\ldots,f_p$ such that
$f=f_1 f_2 \cdots f_p$ and every $f_i$ has degree at least $\ell$.
Denote $\minvar(f) = \min_{i = 1,\ldots,p} n_i$,
where $n_i$ is the number of variables that $f_i$ is defined over.

The following lemma shows that a small constant depth multilinear formula
can be written as a short sum of formed polynomials.

\begin{lem}
\label{lem: structure of const depth}
Let $\Phi$ be a multilinear homogeneous formula of size $s$ and product-depth $d$
computing a polynomial of degree $k$.
Let $q > 1$ be a natural number such that $k (2q)^{-d} > 1$.
Then there exist $(q,k (2q)^{-d})$-form polynomials $f_1,\ldots,f_{s'}$ such that $z_1/k_1 < z_i/k_i$, 
$$\wh{\Phi} = f_1 + \cdots + f_{s'} $$
and $\sum_{i=1,\ldots,s'} \minvar(f_i) \leq s$.
\end{lem}

\begin{proof}
First let us note the following:
\begin{clm}
\label{clm: exist node in const depth}
Let $r > 1$ be a real number such that $kr^{-d} > 1$.
Then there exists a product node $w$ in $\Phi$ such that
$\deg(w) \geq kr^{-d+1}$ and $\deg(v) < \deg(w)/r$ for every child $v$ of $w$.
Moreover, if $r = 2q$ with $q \in \N$, then $\wh{\Phi}_w$ is in $(q, k (2q)^{-d})$-form.
\end{clm}

\begin{proof}
The proof is by induction on $d$.
If $d=1$ and $u=u_1 \times u_2 \cdots \times u_j$ is a product node in $\Phi$,
then $\deg(u) = k$ and $\deg(u_i) \leq 1 < k/r$.
So we can set $w=u$.
Assume that $d>1$, and let $u=u_1 \times u_2 \cdots \times u_j$ be a product node in $\Phi$ with $\deg(u)=k$.
If for every $i=1,\ldots,j$, $\deg(u_i)< k/r$, then we can set $w=u$.
Otherwise there exists $u_i$ such that $\deg(u_i) \geq k/r$.
In this case, $\Phi_{u_i}$ is of product-depth $d' < d$ and degree at least $k/r$.
By the inductive assumption, there exists a product node $w$ in $\Phi_{u_i}$ such that
$\deg(w) \geq \deg(u_i) r^{-d'+1} \geq k r^{-d+1}$ with the desired property.

Let $f$ be a polynomial of degree at least $m$.
If $f=f_1 f_2 \cdots f_n$ with $\deg(f_i) < m/t$, $t \in \N$, for every $i = 1,\ldots,n$,
then $f$ is of $(\lfloor t/2 \rfloor,m/t)$-form;
this is achieved by an appropriate grouping of $f_1,\ldots,f_n$.
Hence if $r=2q$, the node $w$ defines a polynomial of $(q,k (2q)^{-d})$-form.
\end{proof}

Let $w$ be a node given by Claim~\ref{clm: exist node in const depth}.
As in the proof of Lemma~\ref{lem: structure of formulas}, we can write
$$\wh{\Phi} = h \cdot \wh{\Phi}_w + \wh{\Phi}_{(w=0)}.$$
Let $s_w$ denote the size of $\Phi_w$ and let $s_{(w=0)}$ denote the size of $\Phi_{(w=0)}$.
The polynomial $\wh{\Phi}_{(w=0)}$ is either zero or of degree $k$.
In the latter case, by induction, it can be written as
$\sum_{i=1,\ldots,s_{(w=0)}'} g_i$ with $s_{(w=0)}' \leq s_{(w=0)}$,
where the $g_i$'s are in $(q,k(2q)^{-d})$-form
and $\sum_{i=1,\ldots, s_{(w=0)}'} \minvar(g_i) \leq s_{(w=0)}$.
The polynomial $\wh{\Phi}_w$ is in $(q, k(2q)^{-d})$-form.
Moreover, if it is written as $f_1 \cdots f_q$,
then every $f_i$ contains at most $s_{w}$ variables.
Since $q>1$ and by multilinearity,
the polynomial $f=(h f_1) f_2 \cdots f_q$ is a polynomial of $(q,k(2q)^{-d})$-form with $\minvar(f) \leq s_w$.
Altogether, $\wh{\Phi}$ can be written as $f + \sum_{i=1,\ldots,s_{(w=0)}'} g_i$
where $\minvar(f) + \sum_{i=1,\ldots,s_{(w=0)}'} \minvar(g_i) \leq s_w + s_{(w=0)} \leq s$.
\end{proof}

The following lemma bounds the number of monomials in a formed polynomial.

\begin{lem}
\label{lem: number monomials in formed mult}
Let $f$ be a multilinear polynomial of $(p,\ell)$-form of degree $k$ with at most $n$ variables,
where $2k \leq n$ and $p,\ell \geq 2$.
Then the number of monomials that occur in $f$ is at most
$3 k^{3/2} \ell^{-(p-1)/2} {n \choose k} \minvar(f) / n$.
\end{lem}

\begin{proof}
Assume that $f=f_1 \cdots f_p$, where $f_i$ has degree $k_i$ and
$n_i$ variables, assume without loss of generality that $n_p =
\minvar(f)$. Homogeneity implies $k_1 + \cdots + k_p = k$ and
multilinearity implies $n_1 + \cdots + n_p \leq n$
(without loss of generality $n_1 + \cdots + n_p = n$).
Since each $f_i$ is also homogeneous and multilinear, it contains at most
$\binom{n_i}{k_i}$ monomials. Thus, $f$ contains at most ${n_1
\choose k_1} \cdots \binom{n_{p-1}}{k_{p-1}} {n_p \choose k_p}$
monomials, which, by Lemma~\ref{lem: new binomial}, is at most
$3 k^{1/2} (k_1 \cdots k_{p-1})^{-1/2} {n - n_p \choose k - k_p} {n_p \choose k_p}$.
We have
\begin{align*}
  {n - n_p \choose k - k_p} {n_p \choose k_p}
  = \frac{k - k_p + 1}{n - n_p + 1} {n - n_p + 1 \choose k - k_p + 1}
  \frac{n_p}{k_p} {n_p - 1 \choose k_p - 1}
  \leq \frac{k - k_p + 1}{(n - n_p + 1)k_p} {n \choose k} n_p.
\end{align*}
The minimality of $n_p$ implies $n_p\leq n/p$. Hence 
\[\frac{k - k_p + 1}{(n - n_p + 1)k_p}\leq \frac{k}{(n - n_p)k_p}\leq \frac{k }{n(1-1/p)k_p}\leq \frac{k}{n},\]
where the last inequality follows from the assumption $p,k_p\geq 2$. Therefore ${n - n_p \choose k - k_p} {n_p \choose k_p}\leq \frac{k}{n} \binom{n}{k}n_p$
and the lemma follows.
\end{proof}

The following proposition bounds the number of monomials in a polynomial
that has a small multilinear homogeneous formula of constant depth.

\begin{prop}
\label{prop: lower bound const d}
Let $\Phi$ be a multilinear homogeneous formula of size $s$, degree $k$, product-depth $d$,
and over at most $n$ variables, where $n\geq 2k$ and $k^{1/d} \geq 8$.
Then the number of monomials that occur in $\wh{\Phi}$ is at most
$6 k^{3/2} 2^{-k^{1/d}/8} {n \choose k} s / n$.
\end{prop}

\begin{proof}
Let $q = \lfloor k^{1/d}/4 \rfloor \geq 2$ and let $\ell = k (2q)^{-d} \geq 2$.
Combining Lemmas
\ref{lem: number monomials in formed mult} and
\ref{lem: structure of const depth},
the polynomial $\wh{\Phi}$ contains at most
$3 k^{3/2} \ell^{-(q-1)/2} {n \choose k} s / n$.
Since $\ell^{-(q-1)/2} \leq 2 \cdot 2^{-k^{1/d}/8}$, the proposition follows.
\end{proof}

\begin{cor} The second part of Theorem~\ref{thm: lb} holds.
\end{cor}

\begin{proof}
The number of monomials in $S^k_n$ is ${n \choose k}$.
\end{proof}

\section{Upper bounds and separations}
\label{sec: upper bound}

In this section we show several upper bounds on the complexity of the symmetric polynomials.
We consider four models of computation in the following subsections.

\subsection{Multilinear nonhomogeneous depth three}\label{sec: Ben-Or}

We now show that $S^k_n$ can be computed by multilinear formulas of depth three
(and product-depth one) of size $O(n^2)$.
These formulas are of course not homogeneous, and we obtain a separation between homogeneous multilinear and non-homogeneous multilinear formulas.
The construction was first suggested by Ben-Or (see \cite{SW}),
and we give it here for completeness.

For $t \in \R$, denote
$$f_t = (x_1t + 1)(x_2t +1)\cdots (x_nt + 1) = \sum_{k = 0}^n t^k S_{n}^k .$$
Evaluating at $t = 1,\ldots,n+1$,
$$ \left[%
\begin{array}{c}
  f_1 \\
  f_2 \\
  \ldots \\
  f_{n+1} \\
\end{array}%
\right] =
A \
\left[%
\begin{array}{c}
  S^0_{n} \\
  S^1_{n} \\
  \cdots \\
  S^n_{n} \\
\end{array}%
\right] $$
with
$$ A = \left[%
\begin{array}{cccc}
  1^0 & 1^1 & \cdots & 1^{n} \\
  2^0 & 2^1 & \cdots & 2^{n} \\
   &  & \cdots &  \\
  (n+1)^0 & (n+1)^1 & \cdots & (n+1)^n \\
\end{array}%
\right] .$$
Since the matrix $A$ is invertible, we can express every $S^k_{n}$ as a linear combination of $f_1,\ldots,f_{n+1}$.
Since $f_t$ has a formula of depth two and size roughly $n$ computing it,
we can compute the symmetric polynomials with a depth three formula of size roughly $n^2$.
(The same argument holds whenever there are more than $n$ nonzero elements in the underlying field.)

\subsection{Homogeneous non-multilinear}\label{sec:homogeneous upper bound}

We  now give an upper bound on the homogeneous formula size of $S^k_n$.
Let $w$ be a weight function that assigns a positive natural number $w(x)$ to every variable $x$.
The \emph{$w$-degree} of a monomial $x_{i_1} x_{i_2} \cdots x_{i_k}$
is defined as $w(x_{i_1}) + w(x_{i_2}) + \cdots + w(x_{i_k})$.
A constant has $w$-degree zero.
We say that a polynomial $f$  is \emph{$w$-homogeneous} if all monomials in $f$ have the same $w$-degree.
A circuit $\Phi$ is \emph{$w$-homogeneous} if every node in $\Phi$ computes a $w$-homogeneous polynomial.

\begin{lem}
\label{lem: two parts}
\begin{enumerate}

\item \label{itm: one} Let $\Phi$ be a $w$-homogeneous formula in variables $x_1,\ldots,x_k$,
and let $\phi_1,\ldots,\phi_k$ be homogeneous formulas of degrees $w(x_1),\ldots,w(x_k)$.
Then the formula $\Phi(\phi_1,\phi_2,\ldots,\phi_k)$ is homogeneous
of degree that is equal to the $w$-degree of $\Phi$;
the formula $\Phi(\phi_1,\phi_2,\ldots,\phi_k)$ is obtained by substituting the formula $\phi_i$
instead of $x_i$ for every $i = 1,\ldots,k$.

\item \label{itm: two} 
Let $f$ be a polynomial of degree $k$ that has a $w$-homogeneous circuit of size $s$,
then $f$ has a $w$-homogeneous formula of size $(sk)^{O(\log k)}$.

\end{enumerate}
\end{lem}
\begin{proof}
 \ref{itm: one} is by a straightforward induction on the size of $\Phi$.\ignore{
When $\Phi$ is an input node, the lemma holds since each of the $\phi_i$'s are homogeneous.
When the output node of $\Phi$ is a product node,
the lemma holds by induction since the product of homogeneous formulas is homogeneous
and since the $w$-degree of $fg$ is the sum of the $w$-degree of $f$ and the $w$-degree of $g$
for two $w$-homogeneous polynomials $f$ and $g$.
When the output node in a plus node,
the lemma holds by induction since the sum of homogeneous polynomials of the same degree is homogeneous
with the same degree (and since a similar statement holds for $w$-degree as well).}

The proof of \ref{itm: two} follows by the construction in \cite{Hyafil} --
this construction transforms a $w$-homogeneous circuit into a $w$-homogeneous formula
with the appropriate size.
Here is a rough sketch of the construction.
Let $\Phi$ be the circuit computing $f$
(assume without loss of generality that the in-degree of $\Phi$ is at most two).
Let $V$ be the set of nodes $v$ in $\Phi$ such that the $w$-degree of $v$ is at least $k/2$,
and $v= v_1 \times v_2$ with the $w$-degrees of both $v_1$ and $v_2$ less than $k/2$.
It can be shown that $f = \sum_{v \in V} h_v \wh{\Phi}_{v_1} \wh{\Phi}_{v_2}$
with $h_v$ having a circuit of size at most roughly the size of $\Phi$.
If we denote by $L(s,k)$ the smallest formula for a polynomial of degree $k$ that has a circuit of size $s$,
we have that $L(s,k)$ is at most roughly $s L(s,k/2)$.
Thus $L(s,k)$ is at most roughly $s^{\log k}$.
\end{proof}

\begin{thm} $S^k_n$ has a homogeneous formula of size $k^{O(\log k)}n$,
and a depth four homogenous formula of size $2^{O(k^{1/2})}n$.
\end{thm}

\begin{proof}
An application of Newton's identities.
Let $P^k_n$ be the polynomial $\sum_{i=1,\ldots, n}x_i^k$.
Let $Z_k$ be a polynomial in the variables $y_1,\ldots,y_k$ defined inductively
as $Z_0 = 1$, and for $k \geq 0$,
\begin{align*}
Z_{k+1}= \frac{1}{k+1} \big( y_1 \cdot Z_k - y_2\cdot Z_{k-1}+ y_3\cdot Z_{k-2}- \cdots  +(-1)^{k+1} y_{k+1}\cdot Z_0 \big).
\end{align*}
Newton's identities assert that \[S^k_n=Z_k(P^1_n,\ldots,P^k_n).\]
Define the weight $w$ as $w(y_i)=i$.
Thus $Z_k$ is a $w$-homogeneous polynomial of $w$-degree $k$ and degree $k$
(this follows by induction on $k$).
The definition of $Z_k$ shows that it has a $w$-homogeneous circuit of size $O(k^2)$.
By Lemma~\ref{lem: two parts}, there exists a $w$-homogeneous formula of size $k^{O(\log k)}$ computing $Z_k$.
Since the degree of $P^i_n$ is $i$ and it has a homogeneous formula of size $kn$,
the polynomial $S^k_n= Z_k(P^1_n,\ldots,P^k_n)$ has a homogenous formula of size $k^{O(\log k)}n$.

Since $Z_k$ is $w$-homogeneous of $w$-degree $k$, the only monomials that occur in it are of the form
$y_{i_1} y_{i_2} \cdots y_{i_t}$ with $i_1 + i_2 + \cdots + i_t = k$.
The number of $i_1 \geq i_2 \geq \cdots \geq i_t$ that sum up to $k$ is known as the \emph{partition function} of $k$.
A classical result of Hardy and Ramanujan says that the partition function of $k$ is at most $2^{O(k^{1/2})}$.
Thus $Z_k$ has $2^{O(k^{1/2})}$ monomials, and so it has a depth two formula of size $2^{O(k^{1/2})}$,
which implies that $S^k_n$ has a depth four homogeneous formula of the appropriate size.
\end{proof}

\subsection{Monotone}
 Let $L(k,n)$ denote the size of a smallest monotone formula computing $S^k_n$. We present an elementary upper bound on $L(k,n)$. The main features of the estimate are the following:
 \begin{enumerate}

 \item $L(k,n)$ is polynomial, if $k\leq \log n$. Moreover,  $L(\log n, n)= O(n^3)$.

\item $L(k,n)= n^{O(\log(n))}$, if $k\geq	\sqrt{n}$.

 \item $L(k,n)
 =  O(n \log^{k-1}n)$, for a fixed $k$. More exactly, $L(k,n)\leq 3n \left( e \frac{\log n}{k-1}\right)^{k-1}$,
 if $k$ is fixed and $n$ sufficiently large.
 \end{enumerate}
 
 \bigskip
\begin{thm} \label{thm: monotone upper bound} 
If $k\geq 2$ then 
\begin{eqnarray*}
L(k,n)&\leq& 2n\cdot n^{\log\left(\frac{k-1}{\log (2n)}+1\right)}\cdot \left (\frac{\log (2n)}{k-1}+1\right)^{k-1}
\end{eqnarray*}
Hence $L(k,n)$ can be written as  $n^{O(\log(\frac{k}{\log n}))}$.
\end{thm}
\begin{proof}
Let us assume that $n$ is power of two. Otherwise choose $n^\prime$ which is a power of two such that
$n < n^\prime<2n$. 
 Recall that we define formula size as the number of leaves. 
Hence $L(1,n) = n$.
Since
\begin{align*}
S^k_n(x_1,\ldots,x_{2n})
&=\sum_{i=0,\ldots,k} S^i_n(x_1,\ldots,x_{n}) S^{k-i}_n(x_{n+1},\ldots,x_{2n}) ,
\end{align*}
we obtain $L(k,2n)\leq 2 \sum_{i=1,\ldots,k} L(i,n)$. Hence in order to upper bound $L(k,n)$, it is sufficient to find a nonnegative function $g$ s.t.
\begin{equation}\label{eq: mbound}
g(2n,k)\geq 2 \sum_{i=1,\ldots,k} g(i,n), ~~~~~ g(n,1)\geq n,
\end{equation}
for every $n,k\geq 1$

Let us first show the following:

\begin{clm} Let $\alpha>0$ be a fixed paramater. Then $
g(n,k)= \frac{n^{1+\alpha}}{(1-2^{-\alpha})^{k-1}}$
satisfies (\ref{eq: mbound}).
\end{clm}
\begin{proof}
Consider $g(n,k)=n^{1+\alpha}\beta^{k-1}$. Then $g(n,1)\geq 1$ if $n\geq 1$ and $\alpha\geq 0$. In order to satisfy (\ref{eq: mbound}), it suffices to have
\begin{eqnarray*} (2n)^{1+\alpha}\beta^{k-1}&\geq& 2n^{1+\alpha} \beta^{k-1}+ 2n^{1+\alpha}\sum_{i=1,\ldots,k-1} \beta^{i-1}, ~~~~~ \hbox{resp.}\\
\beta^{k-1}&\geq& (2^{\alpha}-1)^{-1}\sum_{i=1,\ldots,k-1} \beta^{i-1}.
\end{eqnarray*}
This holds if $\beta= 1+(2^{\alpha}-1)^{-1}= (1-2^{-\alpha})^{-1}$.
\end{proof}

The claim shows that for every $\alpha>0$, 
 $L(k,n)\leq \frac{n^{1+\alpha}}{(1-2^{-\alpha})^{k-1}}$.
Let $z:= \frac{k-1}{\log n}$ and $\alpha:=\log(1+z)$.
Then
\begin{eqnarray*}
\frac{n^{1+\alpha}}{(1-2^{-\alpha})^{k-1}}
        &= & \frac{n^{1+\alpha}}{(z/(1+z))^{k-1}}\\  
 &=&  n^{1+\log(1+z)}\left(1+z^{-1}\right)^{k-1}.
\end{eqnarray*}
This gives  the statement of the theorem.\end{proof}

\subsubsection*{Weakly equivalent polynomials and Boolean complexity}
\label{sec: weakly equivalent}
We say that two polynomials $f$ and $g$ are \emph{weakly equivalent} if for every monomial $\alpha$,
the coefficient of $\alpha$ is nonzero in $f$ iff its coefficient in $g$ is nonzero.
Results in Boolean complexity yield better upper bounds for a monotone polynomial weakly equivalent to $S^k_n$ 
than the ones in Theorem \ref{thm: monotone upper bound}. 
As shown in \cite{Khasin,Friedman}, $k$-threshold function $\mathrm{Th}^k_n$ 
has monotone Boolean formulas of size $O(n\log n)$, if $k$ is fixed. In fact, the construction gives  a monotone arithmetic formula computing a monotone polynomial weakly equivalent to $S^k_n$. 
Our lower bounds apply to any polynomial weakly equivalent to $S^k_n$.  This
shows that using our techniques we cannot hope to prove better lower bounds than $\Omega(n\log n)$, if $k$ is fixed.

In the converse direction, a monotone arithmetic formula computing $S^k_n$, or a weakly equivalent polynomial,
can be interpreted as a monotone Boolean formula computing $\mathrm{Th}^k_n$. Since for $k\geq 2$ such a formula must be of size $\Omega(n\log n)$ (see \cite{Hansel}), we have $\Omega(n\log n)$ lower bound on the size of monotone formulas computing $S^k_n$, or a weakly equivalent polynomial.

Finally, observe that if $S^k_n$ has a multilinear homogeneous formula $\Phi$ of size $s$, then there exists a monotone formula $\Phi^\prime$ of size $s$ computing a monotone polynomial weakly equivalent to $S^k_n$.
(The formula  $\Phi^\prime$ is obtained by replacing every constant $a$ in $\Phi$ by $\vert a\vert$.)
Hence the lower bound $\Omega(n\log n)$ applies also to homogeneous multilinear formulas computing $S^k_n$, $k\geq 2$. 

\subsection{Noncommutative}
\label{sec: noncom}

A \emph{noncommutative} polynomial over a field $\F$ is a polynomial in which the variables
do not multiplicatively commute, for example, $x_1x_2$ and $x_2 x_1$ are two different polynomials.
A \emph{noncommutative} formula is a formula which we understand as computing a noncommutative polynomial.
Exponential lower bounds on the size of noncommutative formulas computing determinant and  permanent were given in
\cite{Nisan}. In that paper, Nisan posed the problem of separating monotone and general noncommutative formulas.
Let us define $S^k_n$ as the noncommutative polynomial
\[\sum_{ i_1< i_2<\cdots <i_k} x_{i_1} x_{i_2} \cdots x_{i_k}.\]
The lower bound from Section~\ref{sec: formulas} and the upper bound from
Section~\ref{sec: Ben-Or} apply also to noncommutative setting, and yield: 
\begin{prop} $S^n_{2n}$ has a noncommutative formula of size $O(n^2)$,
but every monotone noncommutative formula for it has size at least $n^{O(\log n)}$.
\end{prop}

\section{Summary}

Whereas Boolean complexity of threshold functions has been mapped quite accurately,
the arithmetic complexity of  symmetric polynomials is folded in subtle mist.
Here we summarise the basic known results on the formula complexity of $S^k_n$.

\bigskip
{\it
\begin{tabular}{l c  c}
  & Lower bound & Upper bound \vspace{0.1cm}\\
~~~~~Depth three, infinite fields  \footnotemark[3]  \vspace{0.1cm} & $\Omega(n^2)$, if $k\sim n$  & $O(n^2)$ \\
~~~~~Homogeneous  \vspace{0.1cm} & & $k^{O(\log k)} n$\\
~~~~~Homogeneous multilinear    \vspace{0.1cm}& $k^{\Omega(\log k)}n$ &    $n^{O(\log(\frac{k}{\log n}))}$ \\

~~~~~Homogeneous depth three  \footnotemark[4]\vspace{0.1cm}  &  $\binom{n}{\lfloor k/2\rfloor} 2^{-k}$ &   \\
~~~~~Homogeneous depth four \vspace{0.1cm}&  & $2^{O(k^{1/2})}n$\\
~~~~~Homog. mult. product-depth $d$\vspace{0.1cm}  & $2^{\Omega(k^{1/d})}n$ & \\

\end{tabular}
}

\footnotetext[3]{See \cite{SW}.}
\footnotetext[4]{See \cite{NW}.}

\bigskip
Monotone bounds are the same as the multilinear homogeneous ones, and in both cases we can add the lower bound $\Omega(n\log n)$ taken from monotone Boolean complexity of threshold functions (see Section~\ref{sec: weakly equivalent}). 

Note that  the lower bound and the upper bound on multilinear homogeneous complexity are both polynomial,  if $k= \log n$, both superpolynomial, if  $k= n/2$, but if $k=\log^2 n$, the lower bound is polynomial, whereas the upper bound is $n^{O(\log\log n)}$. The
 `match' between multilinear homogeneous lower bounds and homogeneous upper bounds is also slightly irritating. However, the bounds cannot be exactly the same, for in the multilienear homogeneous case, we need at least $\Omega(n\log n)$ if $k\geq 2$.

Let us end with the following two questions:
{\it
\begin{enumerate}\item
Can $S^k_n$ be computed by a monotone formula of size $\mathrm{poly(n)}\cdot k^{O(\log k)}$?
\item Does the central symmetric polynomial $S^n_{2n}$ have polynomial size homogeneous formula? 
\end{enumerate}
}

\bibliographystyle{plain}

\end{document}